\documentclass[twocolumn]{aastex631}

\usepackage{bm}
\shorttitle{AASTeX v6.3.1 Sample article}
\shortauthors{Steiger et al.}
\graphicspath{{./}{figures/}}
\begin{document}

\title{Probing Photon Statistics in Adaptive Optics Images with SCExAO/MEC\footnote{Based in part on data collected at Subaru Telescope, which is operated by the National Astronomical Observatory of Japan.}}

\author[0000-0002-4787-3285]{Sarah Steiger}
\correspondingauthor{Sarah Steiger}
\email{steiger@physics.ucsb.edu}
\affiliation{Department of Physics, University of California, Santa Barbara, Santa Barbara, California, USA}

\author{Timothy D. Brandt}
\affiliation{Department of Physics, University of California, Santa Barbara, Santa Barbara, California, USA}

\author{Olivier Guyon}
\affiliation{Subaru Telescope, National Astronomical Observatory of Japan, 
650 North A`oh$\bar{o}$k$\bar{u}$ Place, Hilo, HI  96720, USA}
\affil{Steward Observatory, The University of Arizona, Tucson, AZ 85721, USA}
\affil{College of Optical Sciences, University of Arizona, Tucson, AZ 85721, USA}
\affil{Astrobiology Center, 2-21-1, Osawa, Mitaka, Tokyo, 181-8588, Japan}

\author[0000-0001-5721-8973]{Noah Swimmer}
\affiliation{Department of Physics, University of California, Santa Barbara, Santa Barbara, California, USA}

\author{Alexander B. Walter}
\affiliation{Jet Propulsion Laboratory, California Institute of Technology, Pasadena, California 91125, USA}

\author{Clinton Bockstiegel}
\affiliation{IQM, Keilaranta 19, 02150 Espoo, Finland}

\author{Julien Lozi}
\affiliation{Subaru Telescope, National Astronomical Observatory of Japan, 
650 North A`oh$\bar{o}$k$\bar{u}$ Place, Hilo, HI  96720, USA}

\author[0000-0003-4514-7906]{Vincent Deo}
\affiliation{Subaru Telescope, National Astronomical Observatory of Japan, 
650 North A`oh$\bar{o}$k$\bar{u}$ Place, Hilo, HI  96720, USA}

\author{Sebastien Vievard}
\affiliation{Subaru Telescope, National Astronomical Observatory of Japan, 
650 North A`oh$\bar{o}$k$\bar{u}$ Place, Hilo, HI  96720, USA}

\author[0000-0002-9372-5056]{Nour Skaf}
\affiliation{Subaru Telescope, National Astronomical Observatory of Japan, 
650 North A`oh$\bar{o}$k$\bar{u}$ Place, Hilo, HI  96720, USA}
\affiliation{LESIA, Observatoire de Paris, Univ.~PSL, CNRS, Sorbonne Univ., Univ.~de Paris, 5 pl. Jules Janssen, 92195 Meudon, France}
\affiliation{Department of Physics and Astronomy, University College London, London, United Kingdom}

\author{Kyohoon Ahn}
\affiliation{Subaru Telescope, National Astronomical Observatory of Japan, 
650 North A`oh$\bar{o}$k$\bar{u}$ Place, Hilo, HI  96720, USA}

\author[0000-0001-5213-6207]{Nemanja Jovanovic}
\affiliation{Department of Astronomy, California Institute of Technology, 1200 E. California Blvd.,Pasadena, CA, 91125, USA}

\author{Frantz Martinache}
\affiliation{Universit\'{e} C\^{o}te d'Azur, Observatoire de la C\^{o}te d'Azur, CNRS, Laboratoire Lagrange, France}

\author[0000-0003-0526-1114]{Benjamin A. Mazin}
\affiliation{Department of Physics, University of California, Santa Barbara, Santa Barbara, California, USA}






\begin{abstract}
    We present an experimental study of photon statistics for high-contrast imaging with the Microwave Kinetic Inductance Detector (MKID) Exoplanet Camera (MEC) located behind the Subaru Coronagraphic Extreme Adaptive Optics System (SCExAO) at the Subaru Telescope. We show that MEC measures the expected distributions for both on-axis companion intensity and off-axis intensity which manifests as quasi-static speckles in the image plane and currently limits high-contrast imaging performance. These statistics can be probed by any MEC observation due to the photon-counting capabilities of MKID detectors. Photon arrival time statistics can also be used to directly distinguish companions from speckles using a post-processing technique called Stochastic Speckle Discrimination (SSD). Here, we we give an overview of the SSD technique and highlight the first demonstration of SSD on an extended source \textemdash the protoplanetary disk AB Aurigae. We then present simulations that provide an in-depth exploration as to the current limitations of an extension of the SSD technique called Photon-Counting SSD (PCSSD) to provide a path forward for transitioning PCSSD from simulations to on-sky results. We end with a discussion of how to further improve the efficacy of such arrival time based post-processing techniques applicable to both MKIDs, as well as other high speed astronomical cameras.  
\end{abstract}

\section{Introduction}
Direct imaging is an extremely technologically challenging exoplanet detection technique with necessary contrasts easily exceeding $10^{-6}$ for even the largest self-luminous planets with orbits wider than that of Saturn ($>$ 10 au). For this reason, nearly all of the $\sim$ 10--20 directly imaged planets have separations of 10--150 au, typically $\rho$ $\sim$ 0\farcs{}4--2\arcsec{} on the sky \citep[e.g.][]{Marois2008a,Lagrange2009, Rameau2013, kuzuhara2013, Currie2014a,macintosh2015,Chauvin2017}. In order to directly image a new regime of planets at Jupiter-to-Saturn like separations, new technologies and techniques need to be developed to push for higher contrasts at smaller inner working angles (IWA).

The current limitation for ground-based direct imaging is point spread function (PSF) sized ``speckles'' which are caused by uncontrolled diffracted and scattered starlight and have a range of correlation timescales ($\tau$) and sources. Rapidly-evolving atmospheric speckles ($\tau$ $\sim$ 1-20 ms) result from aberrations left uncorrected by an adaptive optics (AO) system and average out over the course of long-exposure images, forming a smooth halo \citep[e.g.][]{Perrin2003,Soummer2007}. These ``fast" speckles are easier to remove and can be corrected by improved AO control loops which will mitigate temporal bandwidth error and measurement (photon noise) error \citep[e.g.][]{Guyon2005}. Alternatively, quasi-static speckles result from instrument imperfections such as non-common path errors, telescope vibrations, etc. \citep{Guyon2005, lozi2018}. These speckles do not average out over long exposures, can interfere with atmospheric speckles, and be pinned to the diffraction rings, easily masquerading as point sources in an image \citep{Soummer2007}. This quasi-static speckle noise is temporally well correlated ($\tau$ $\sim$ 10-60 minutes), presenting a fundamental obstacle in exoplanet direct imaging \citep[e.g.][]{Marois2008b}. 

While focal-plane wavefront control methods offer a pathway to suppress these speckles on-sky \citep[e.g.][]{Give'on2007, martinache2016}, they are most commonly removed in post-processing. Unfortunately, the most common post-processing techniques that utilize advanced PSF subtraction methods \citep[e.g.][]{Lafreniere2007,Soummer2012} become less effective at small IWAs where direct detections are most challenging.

Angular Differential Imaging \citep[ADI;][]{Marois2006} uses the rotation of the Earth or, analogously, the field of view rotation of an altitude-azimuth telescope with the field derotator turned off, to separate speckles (which will remain stationary) from planetary companions (which will rotate with the sky). The arc traced by an object across the image, however, scales proportionally with angular separation for a given unit time, resulting in less rotation at smaller IWAs. Additionally, the rotation in $\lambda$/D units is smaller within a few diffraction beamwidths, resulting in severe self-subtraction and partial-subtraction of a planet signal at these small separations \citep{Lafreniere2007, mawet2012}. 
Spectral Differential Imaging \citep[SDI;][]{Marois2000} utilizes differences in the wavelength dependent scaling of the diffracted light speckles and companions in polychromatic images, but requires broad spectral coverage close to the primary otherwise it also suffers from self-subtraction effects. Reference Star Differential Imaging \citep[RDI/RSDI;][]{Soummer2012} does not inherently suffer at small IWAs, but requires careful matching between the target of interest and the reference star for optimal performance \citep{ruane2019}. A method to suppress quasi-static speckles that is free of the limitations of ADI, SDI, and RDI would significantly improve our ability to detect jovian planets at Jupiter-to-Saturn like separations.

Post-processing techniques that take advantage of photon arrival time statistics, such as Stochastic Speckle Discrimination \citep[SSD;][]{gladysz2008}, currently show much promise as they use differences in the intensity distributions of speckles and companions at millisecond timescales to distinguish the two populations. Since this technique uses only temporal information in the form of the `instantaneous' intensity, it requires no PSF reference subtraction and is free of the spectral, spatial, and reference matching issues of SDI, ADI, and RDI respectively. Effectively utilizing this information however can be challenging as you need many photons to arrive within a single speckle decorrelation time ($\tau \sim$ 0.1 s). Additionally, cameras with a frame rate faster than this timescale and very low read noise at these high speeds are needed.

Microwave Kinetic Inductance Detector (MKID) instruments are therefore very appealing for these techniques as they are photon-counting instruments with $\mu$s precision and no read noise or dark current (see \citet{mazin2012superconducting, szypryt2017large} for more details on MKIDs). Specifically, the MKID Exoplanet Camera \citep[MEC;][]{walter2019} is a y-J band MKID IFU located behind the Subaru Coronagraphic Extreme Adaptive Optics System \citep[SCExAO;][]{Jovanovic2015} at the Subaru Telescope on Maunakea that has been recently commissioned and has shown its ability to use the temporal resolution afforded by its MKID detector to enable the discovery of low mass companions using SSD \citep{steiger2021}.

In this work, we demonstrate the ability of MKIDs to accurately probe the distinct intensity distributions of quasi-static speckles and companions using data from SCExAO/MEC. We also highlight more generally how others can use MEC to access photon arrival time information and use the techniques discussed here, or to develop new time-based analysis techniques in the future. We then give an overview of the status of current photon statistical post-processing techniques for high-contrast imaging with on-sky MEC SSD results shown for both point sources and diffuse sources. Next we describe a new post-processing technique first shown in \citet{walter2019} called Photon-Counting SSD (PCSSD). Using both simulations and on-sky data, we identify key features that have prevented this technique from achieving desired performance on-sky and the steps that can be taken to improve performance in the future. Finally we end by discussing key paths forward towards developing new and more effective photon arrival time based post-processing techniques for exoplanet direct imaging.  

\section{Photon Statistics in Millisecond AO Images}
Before discussing how to utilize photon arrival time information with MEC and photon arrival time based post-processing techniques, we present an overview of the statistics that govern the different parts of an image downstream of an AO system at millisecond frame rates. 

\subsection{Speckles}
Originally derived by \cite{goodman1975statistical} and experimentally verified by \cite{cagigal2001} and \cite{fitzgerald2006speckle}, the underlying probability density function (PDF) that estimates the intensity distribution of off-axis stellar speckles in the image plane is given by a modified Rician (MR)

\begin{equation}
    p_{MR}(I) = \frac{1}{I_S}\exp\left({-\frac{I + I_C}{I_S}}\right) I_0 \left(\frac{2\sqrt{II_C}}{I_S}\right) 
\end{equation}

where $I_0(x)$ denotes the zero-order modified Bessel function of the first kind, $I_C$ describes the coherent intensity component attributed to the unaberrated PSF of the primary,  and $I_S$ is the time variable component of the total intensity that describes the speckle field \citep[see also][]{Marois2008b}. It is important to note that the shape of this distribution is always positively skewed (i.e. the distribution tail falls to the right hand side of the mean) and that the skewness of the MR can be parameterized by the ratio of $I_C/I_S$ (the larger the $I_C/I_S$ ratio, the less skewed the distribution). 

For a sequence of exposures shorter than the decorrelation time of atmospheric speckles ($\sim$ 10-100 ms), a histogram of the image plane intensity follows a MR and $I_C$ and $I_S$ can be determined for each pixel in an image \citep{fitzgerald2006speckle}. A key feature of MEC is that it stores the arrival time information of every photon and so all time binning can be done in post-processing. This is important since the bin size that ideally samples the MR is difficult to determine a priori and may also vary spatially across the image.

The optimal bin size should be shorter than the decorrelation timescale of the speckles in the image but long enough that each bin contains many photons. If too large of a bin size is chosen, many realizations of the speckle intensity will be averaged over. Conversely, if too small of a bin size is selected, then not enough photons will arrive per bin and the distribution will tend towards Poisson statistics.

\subsection{Companions}

On-axis sources (non-diffracted light, i.e. astronomical objects in the image) behind an Extreme AO (ExAO) system do not follow MR statistics because of the ExAO system itself which can be thought of as a high-pass spatial filter acting on the phase \citep{ Sivaramakrishnan2001}. At the center of an image, the complex amplitude is simply the integral of the pupil complex amplitude as given by Equation \ref{eq:center_phase} \citep{Soummer2007}. 

\begin{equation}
    \Psi\left(0\right) = \int P\left(x\right)e^{i\phi\left(x\right)}dx 
    \label{eq:center_phase}
\end{equation}

Here $P(x)$ denotes the pupil function and $\phi\left(x\right)$ is the phase of the wavefront in the pupil plane. 

With no (or low) AO correction, $\phi\left(x\right)$ is large and so the phase vectors, $e^{i\phi\left(x\right)}$, can take any orientation in the complex plane. Summing a large number of these vectors will produce a random walk resulting in circular Gaussian statistics by the Central Limit Theorem. This adherence to circular Gaussian statistics in phase will result in a MR distribution in intensity and be indistinguishable from off-axis speckles. 

\begin{figure}
    \centering
    \includegraphics[width=\columnwidth]{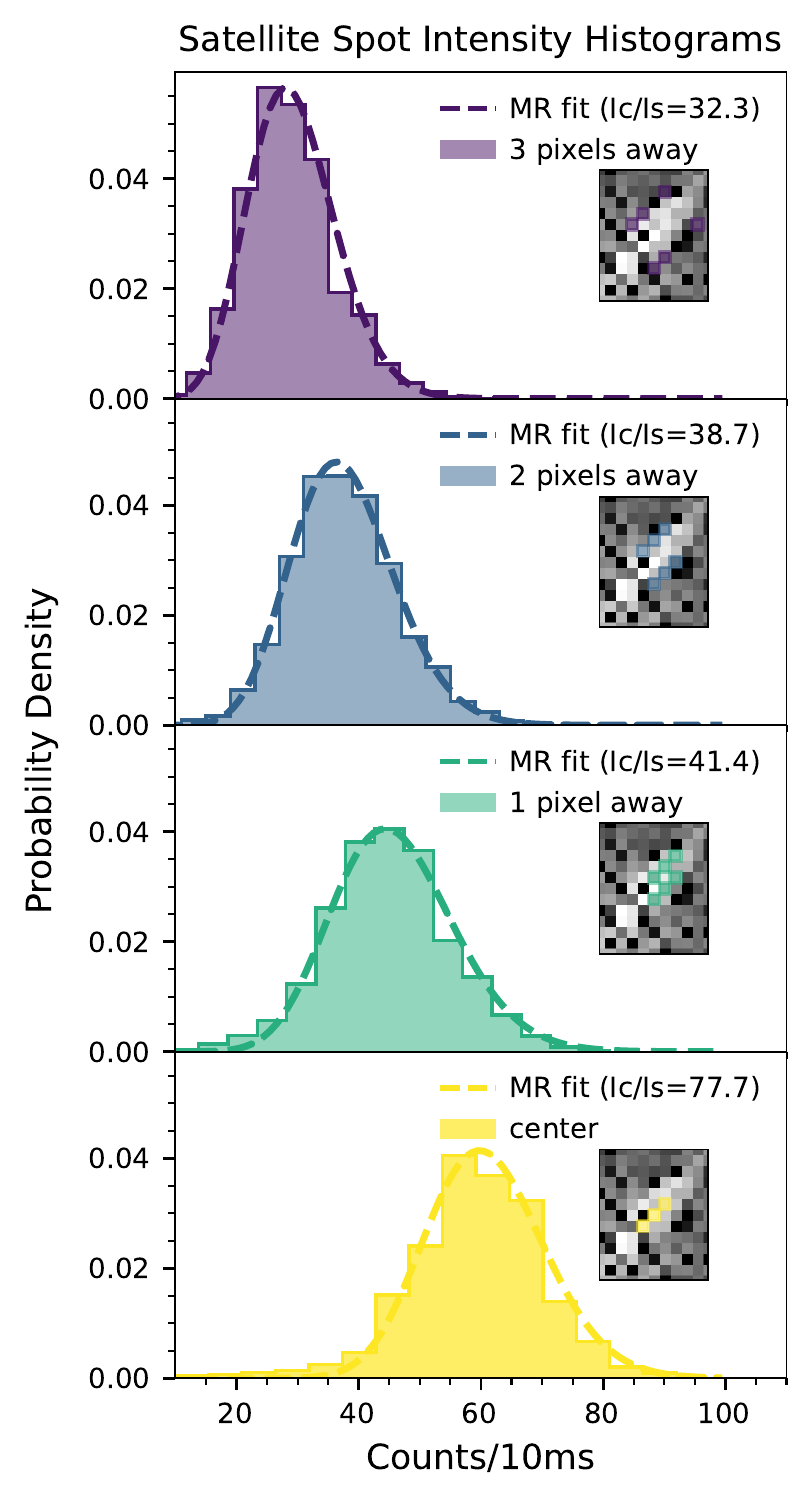}
    \caption{Histograms of the binned satellite spot intensity moving from three pixels away from the center of the spot (purple, top) to the center (yellow, bottom). Histograms shift from positive to negative skewness as they approach the center of the satellite spot as is shown by the best fit MR Ic/Is ratio (dashed lines). Here a higher value indicates less skew. Since this data was taken in y+J band, the satellite spots are elongated in the image and the plotted intensities were found by adding the intensities for six pixels at the same specified distance from the spot center (three on each side). Insets: satellite spot image where the pixels used to generate the histograms are color-coded by their distance from the spot center. Dead pixels were purposefully avoided. }
    \label{fig:sat_hist}
\end{figure}

At high correction levels (i.e.~behind an ExAO system), $\phi\left(x\right)$ is small and so the vectors are not oriented randomly in the complex plane. Summing them will not produce a random walk, the corresponding distribution is not a circular Gaussian, and the resulting on-axis intensity distribution will not follow a MR. The spatial extent where the transition between the on-axis and off-axis intensity distributions occurs is at $\ll 1\lambda/D$ \citep{Soummer2007} which can be seen qualitatively in Figure \ref{fig:sat_hist} -- see also $\S$ \ref{sec:sat_spot}. For reference, at Subaru $1\lambda/D$ is equal to 27.6 mas (2.75 MEC pixels) at 1.1 $\mu$m.  

\begin{figure*}[ht]
    \centering
    \includegraphics[width=0.75\textwidth]{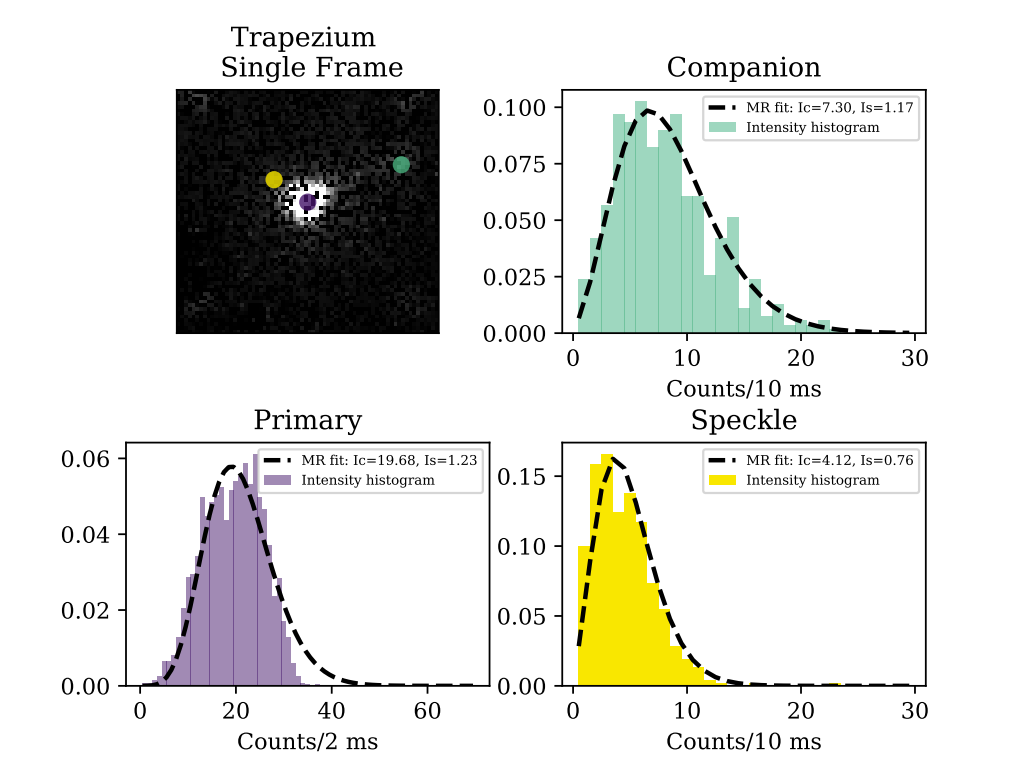}
    \caption{Single-pixel arrival time histograms taken on-sky with SCExAO/MEC of the unocculted star $\Theta$ 1 Orionis (purple, bottom left), a faint secondary companion (teal, upper right), and at a random location in the field with comparable brightness to the faint secondary companion (yellow, bottom right). The best fit MR was plotted for each distribution and is shown with the dashed line. Note that due to the brightness of the primary, 2 ms time bins were used to generate that intensity histogram instead of the 10 ms time bins used for the secondary and field locations. It is clear to see that the MR adequately describes the field location, but breaks down at the location of the primary. The companion is seen to be a convolution of the primary and field PDFs.}
    \label{fig:trap_hists}
\end{figure*}

Instead of following the MR, at these high correction levels the Strehl Ratio (SR) distribution (which is proportional to the intensity) instead follows the PDF described by Equation \ref{eq:gammapdf} which was derived independently by \citet{Soummer2007} and \citet{gladysz2008} :
\begin{equation}
    p_{SR}(sr) = \frac{p_{\hat{\sigma}^2}\left(-\ln\left(sr\right)\right)}{sr}
    \label{eq:gammapdf}
\end{equation}
where $sr$ is the instantaneous SR, and $p_{\hat{\sigma}^2}$ is given by
\begin{equation}
    p\left(x; k, \theta, \mu\right) = \frac{\left(\frac{x-\mu}{\theta}\right)^{k-1} \exp\left(-\frac{x - \mu}{\theta}\right)}{\Gamma\left(k\right)\theta}
    \label{eq:gamma}
\end{equation}
Here $k > 0$ is the shape parameter, $\theta > 0$ is the scale parameter, and $\mu$ is the location parameter, which
shifts the PDF left and right. $\Gamma\left(k\right)$ is a Gamma function. For ease of discussion we will refer to the entirety of Equation \ref{eq:gammapdf} as the `Gamma' distribution for the remainder of this work. 

In contrast to the MR distribution, this Gamma distribution is negatively skewed (i.e. the distribution tail falls to the left hand side of the mean). Figure \ref{fig:trap_hists} illustrates these differences in skewness for various on and off-axis sources with a SCExAO/MEC observation of $\Theta$ 1 Orionis (The Trapezium Cluster). 

\vspace{-0.5cm}
\subsection{Satellite Spots}\label{sec:sat_spot}

Astrometric and spectrophotometric calibrations are very difficult when the target star (typically the only reference in the field-of-view) is obscured by a coronagraph. In order to perform these calibrations, faint copies of the obscured stellar PSF called satellite spots may be intentionally placed into the image plane. For SCExAO, these are generated by placing a ``waffle'' pattern (two orthogonal sine waves) on the AO system's deformable mirror (DM). 

In SCExAO, the satellite spots are additionally modulated at a rate of 2 kHz by flipping the sign of the two sine waves (equivalent to phase shifting them by $\pi$). This is done to make the speckles incoherent with the underlying background and improve photometric performance \citep{Jovanovic2015-astrogrids}. If there is a coherent speckle underneath the spots then these two polarities will not have the same brightness as one will interact constructively and the other destructively. In the regime of these millisecond images, this phase switching is unlikely to affect the statistics at the spot locations as 10s - 100s of these transitions are being averaged over in a single time bin. It is then expected that the satellite spots should follow the same on-axis statistics of the primary.     

Figure \ref{fig:sat_hist} shows intensity histograms moving from the speckle field towards the satellite spot center for a single 25\,s MEC observation of HIP~109427. The statistics shift from positively to negatively skewed, showing that the satellite spot statistics are not only distinct from that of the speckle field, but also qualitatively follow the same distribution expected for the primary. Additionally, this transition between the on and off-axis distributions occurs at the expected location of $\ll 1\lambda/D$ ($< 2.75$ MEC pixels).

The satellite spot statistics themselves are important because if the distribution of the satellite spots matches that of the on-axis source, then they can be used to help fit the free parameters of the on-axis Gamma distribution (Equation \ref{eq:gammapdf}) when observing with a coronagraph. Since the brightness of the satellite spots can be controlled by changing the amplitude of the sine waves on the DM, this would be the simplest way to determine the shape of the companion distribution using on-sky data at high signal-to-noise. Once the shape of this distribution is known it could then enable the use of post-processing techniques that try to explicitly separate the companion and speckle PDFs such as PDF deconvolution (see \cite{gladysz2010}).  

\vspace{2.0cm}
\section{Utilizing Photon Arrival Time Information with MEC}

\begin{figure}
    \centering
    \includegraphics[width=1.05\columnwidth]{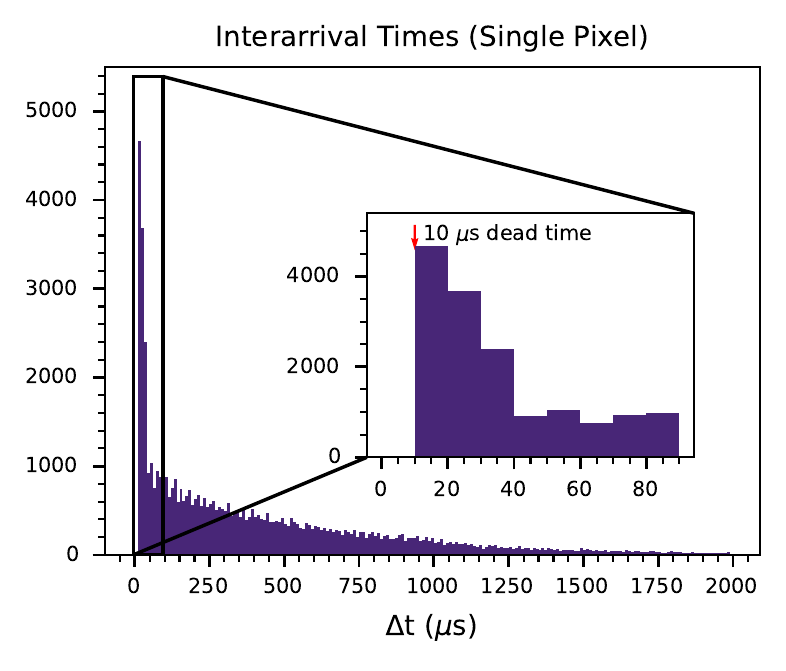}
    \caption{Interarrival time histogram for a single pixel. An excess of interarrival times between 10-40 $\mu$s can be clearly seen that are inconsistent with Poisson statistics.}
    \label{fig:dts_hist}
\end{figure}

Arrival time information can be easily accessed using The MKID Pipeline\footnote{https://github.com/MazinLab/MKIDPipeline} which is open-source and can be downloaded via GitHub \citep{steiger2022}. Since MKID detectors record the arrival time and energy of each incident photon, the format of raw MKID data is a time and energy-tagged photon list that can be queried using the \texttt{Photontable} class on pixel location, time range, photon wavelength, or any combination thereof. A result of this is that all spectral and temporal binning is done in post-processing and MEC has no set `exposure time' for its observations. This is especially beneficial for post-processing techniques that leverage differences in arrival time statistics, like those that are discussed in the following section, as many different timescales can be probed from a single observation. 

\begin{figure*}[ht]
    \centering
    \includegraphics[width=0.7\textwidth]{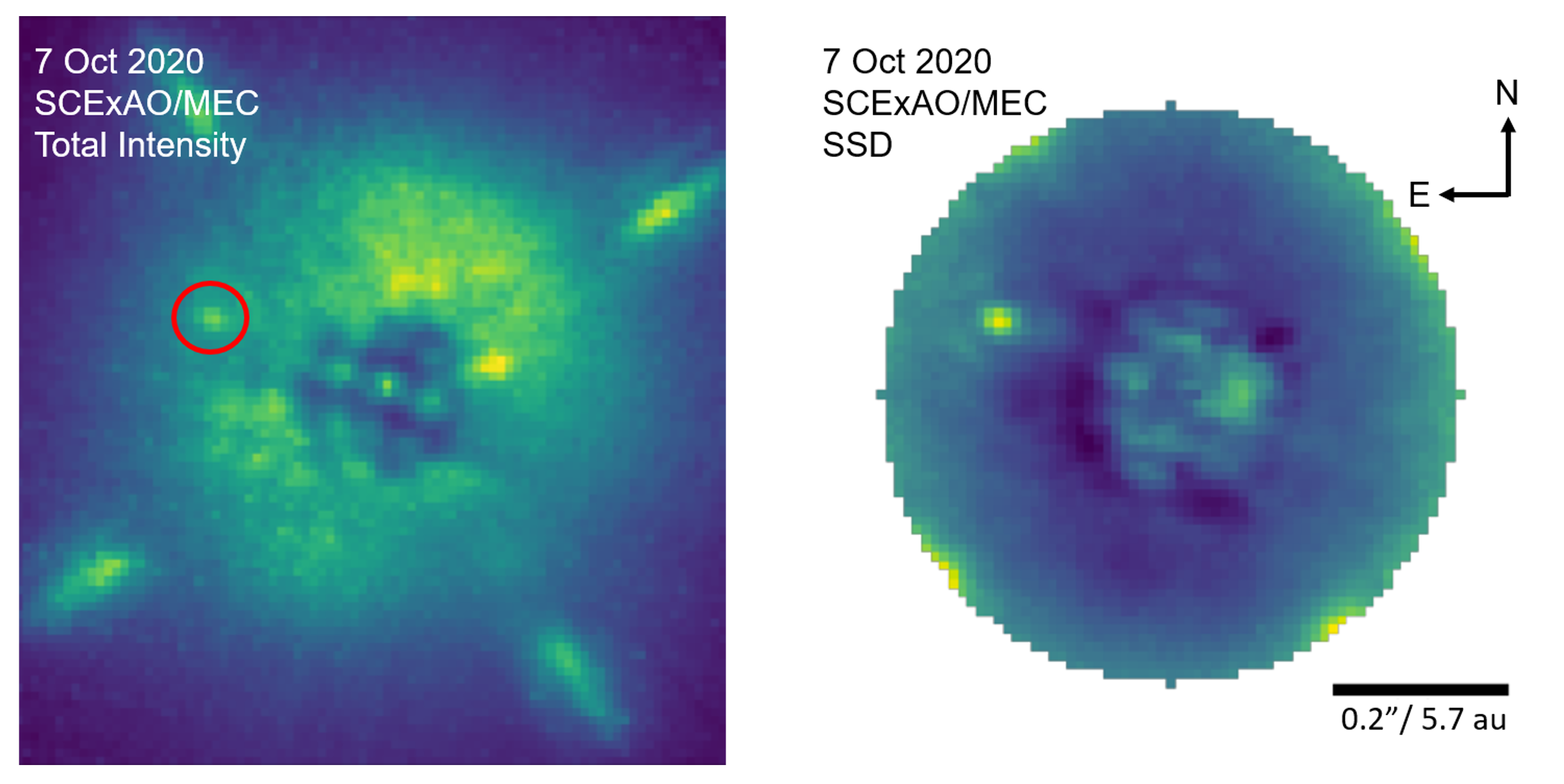}
    \caption{Left: Total intensity image of HIP~109427 B taken with SCExAO/MEC at y and J band where the location of the companion has been circled in red. Right: SSD $I_C$/$I_S$ image of HIP~109427 B. Here the companion is plainly visible as well as dark regions at the edge of the coronagraph showing the removal of pinned speckles from the total intensity image. This Figure is reproduced from \citet{steiger2021}}
    \label{fig:binnedssd}
\end{figure*}

MKIDs temporal resolution limit is determined by the readout speed of the detector ($\sim$1 MHz) and the firmware-imposed dead time. This dead time is set by material properties of the MKID array and for MEC has a value of 10 $\mu$s. During this time, no additional photons are able to be recorded for that pixel to allow it time to return to its idle state \citep{fruitwala2020}.

Immediately after this dead time, a pile-up of photon events has been observed which is likely insignificant for total integrated observations, but can cause unintended effects when using photon arrival time information directly. Empirically this effect decays rapidly after 40 $\mu$s (Figure \ref{fig:dts_hist}) and any work done with MEC that uses arrival time information should take this into account so as to not contaminate results.

\section{Photon Arrival Time  Based Post-Processing Techniques}
\subsection{Stochastic Speckle Discrimination}

Stochastic Speckle Discrimination (SSD) is a post-processing technique first demonstrated by \citet{gladysz2008ssd} that relies solely on photon arrival time statistics to distinguish between speckles and faint companions in coronagraphic images.

It is achieved with MEC by fitting a MR to every pixel in an image with a user-specified temporal bin size. While the components of the MR distribution do not inherently describe the signal from a faint companion, the \textit{ratio} of the coherent component to time variable component - $I_C$/$I_S$ - may reveal faint companions from a comparably bright speckle field \citep{gladysz2009, meeker2018darkness, steiger2021}. This is because the addition of light from a companion (whose statistics follow a negatively skewed Gamma distribution) will make the best-fit MR more negatively skewed at that location. This is analogous to increasing the best fit $I_C$ and results in a larger $I_C$/$I_S$ ratio at the location of the companion compared to other pixels at the same angular separation from the primary. 

SSD was recently shown to be effective on-sky with MEC and facilitated the discovery of a low mass ($\sim$ 0.25 $M_{\odot}$) stellar companion to the nearby A star HIP~109427 (tet Peg), see Figure \ref{fig:binnedssd} and \cite{steiger2021} for more details. 

\subsubsection{SSD on AB Aur}

\begin{figure*}
    \centering
    \includegraphics[width=0.75\textwidth]{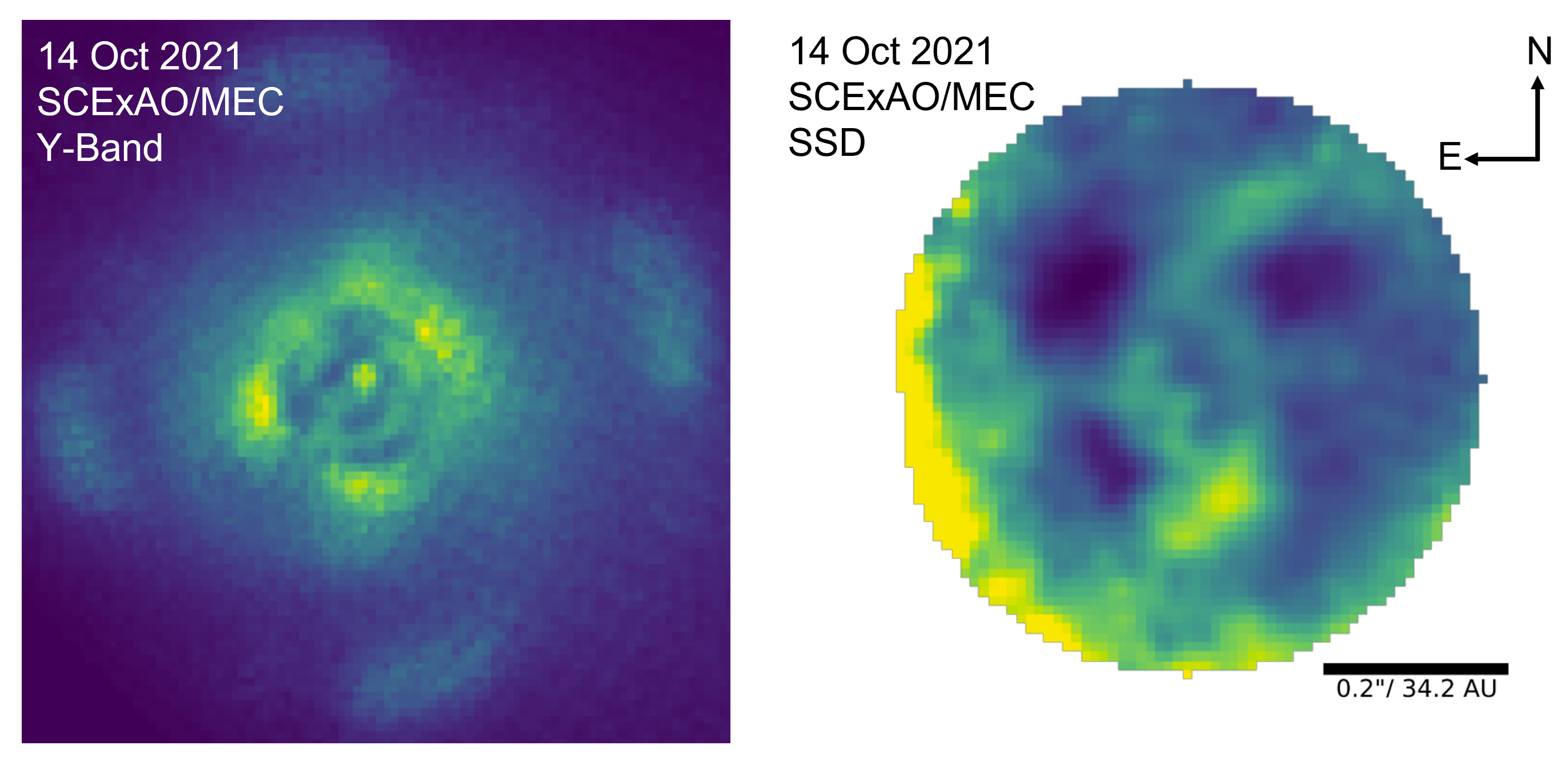}
    \caption{Left: Total y-band intensity dither combined image of AB Aurigae taken with SCExAO/MEC. Disk features here are largely obscured. The satellite spots can be seen as radially smeared bright patches on the edge of this image due to the corrected sky rotation. Right: $I_C$/$I_S$ image of the disk around AB Aurigae clearly showing some of the inner disk features. These include the two main spirals that have roots to the North and South as well as the extended point-like source to the South at $\rho$ = 0\farcs{}16 as found in \cite{Boccaletti2020}. A Gaussian filter has been applied over this image to smooth over small scale inter-pixel variations.}
    \label{fig:AB_aur}
\end{figure*}

Some post-processing techniques, such as ADI, struggle to reveal structures with azimuthal symmetry. SSD does not suffer the same limitations for these sources because the light from the disk will still be on-axis (and thus follow a Gamma distribution) even if it is spread over a region instead of contained within a single PSF. 

Here we tested the performance of SSD on extended sources using a SCExAO/MEC y-band observation of AB Aurigae (AB Aur) taken on 14 October 2021. AB Aur is a well studied system with a known protoplanetary disk that also harbors a recently discovered protoplanet \citep{Boccaletti2020, Currie2022, zhou2022}. AB Aur was observed using SCExAO/MEC for 16 minutes in exceptional seeing conditions of $\sim$0.$\!\!''$3. The results from the SSD reduction of the protoplanetary disk surrounding AB Aur can be seen in Figure \ref{fig:AB_aur} where inner disk structures are revealed in the $I_C/I_S$ map (right) not seen in the y-band total intensity image (left). This bears a strong resemblance to images taken of this system in polarized intensity by the Spectro-Polarimetric High-contrast Exoplanet REsearch instrument at the VLT \citep[SPHERE;][]{beuzit2019} as described in \cite{Boccaletti2020} -- see their Figure 4. In contrast with the SPHERE observations, here only the millisecond intensity distributions were used to generate these images with no polarization information or PSF subtraction techniques employed. It is important to note that this result was in large part facilitated by the exceptional seeing conditions since better seeing leads to a less intense speckle halo and allows the disk's Gamma distributed intensity to significantly modulate the underlying MR distribution of the speckles.  

\subsection{Photon-Counting SSD (PCSSD)}

PCSSD is an extension of the SSD formalism where contributions from an incoherent source of constant intensity ($I_P$) are accounted for in addition to $I_C$ and $I_S$ which define the shape of the MR. Given a list of photon inter-arrival times, the maximum likelihood value of $I_C$, $I_S$ and $I_P$ are determined. Since all inter-photon arrival times are used in this technique, no binning is done and it has been shown to perform twice as well as perfect PSF subtraction on simulated data where the companions were modeled as constant, incoherent sources \citep{walter2019}.

One of the main motivations for expanding the SSD formalism is that the SSD Ic/Is maps -- while helpful for extracting companion astrometry and disk morphology -- only quantify the skewness of the fit MR to the data. This output is therefore not easily converted to physically meaningful units and the images are unhelpful for performing spectroscopy or photometry as would typically be desired to determine key companion properties such as temperature, composition, and mass. PCSSD attempts to solve for this by calculating a likelihood for each inter-photon arrival time so that the output units can be reported in counts or counts/s. Additionally, PCSSD is able to leverage the photon counting nature of MKID detectors to beat the long exposure noise limit by not temporally binning. 

\subsubsection{PCSSD on HIP~109427 B}
%PCSSD has been shown on simulated data to perform up to twice as well as perfect PSF subtraction for extracting companion flux from coronagraphic images -- see \citet{walter2019} for more details. 

In the form described by \cite{walter2019}, PCSSD makes the following assumptions:

\begin{enumerate}
    \item The speckle halo intensity is entirely described by the MR distribution.
    \item $I_C$, $I_S$, and $I_P$ (the intensity of a companion) remain constant over an observation.
    \item Chromaticity is ignored.
\end{enumerate}

The assumptions that are perhaps the most problematic are that the off-axis intensity is entirely described by the MR distribution and that the companion intensity ($I_P$) remains constant over the course of an observation.  MEC has a known infrared (IR) background that can cause the intensity at any pixel to not be fully described by the MR PDF. Additionally, for realistic observing conditions the on-axis companion intensity (which is proportional to the SR) varies quite considerably and we know is described by the Gamma PDF in Equation \ref{eq:gammapdf}.

\begin{figure*}[ht]
    \centering
    \includegraphics[width=0.9\textwidth]{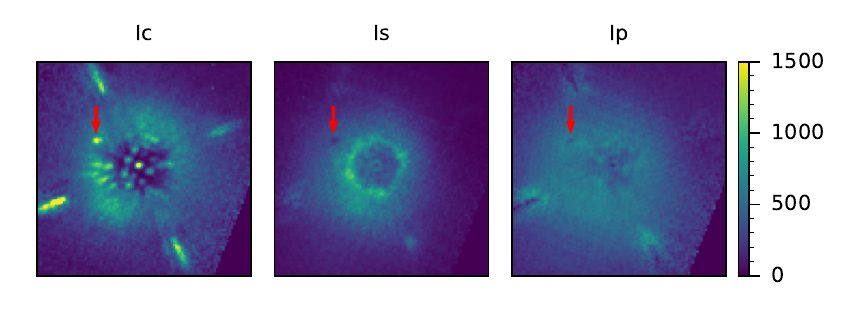}
    \caption{Output of PCSSD on HIP~109427 B (see also Figure \ref{fig:binnedssd}). The red arrows in each image point to the same location of the low mass stellar companion. Here the companion light clearly shows up in $I_C$ with corresponding minima in the intensity in the $I_P$ and $I_S$ images}
    \label{fig:binfree_drizzles}
\end{figure*}

The fact that the on-axis light follows this negatively skewed distribution is in fact one of the main bases that allows traditional SSD to work. Companions stand out in $I_C/I_S$ images due to the addition of the negatively skewed companion PDF which shifts the whole distribution or, analogously, increases the best-fit MR $I_C$ \citep{gladysz2008ssd, meeker2018darkness, steiger2021}.

For this reason, using the PCSSD technique on real data does not accurately separate the companion light into $I_P$, but instead attributes that light largely to $I_C$. See Figure \ref{fig:binfree_drizzles} for an example of PCSSD run on the same HIP~109427 B data set as Figure \ref{fig:binnedssd}. 

\subsubsection{PCSSD on More Realistic Simulated Data}\label{sec:realsimpcssd}

\begin{figure}
    \centering
    \includegraphics{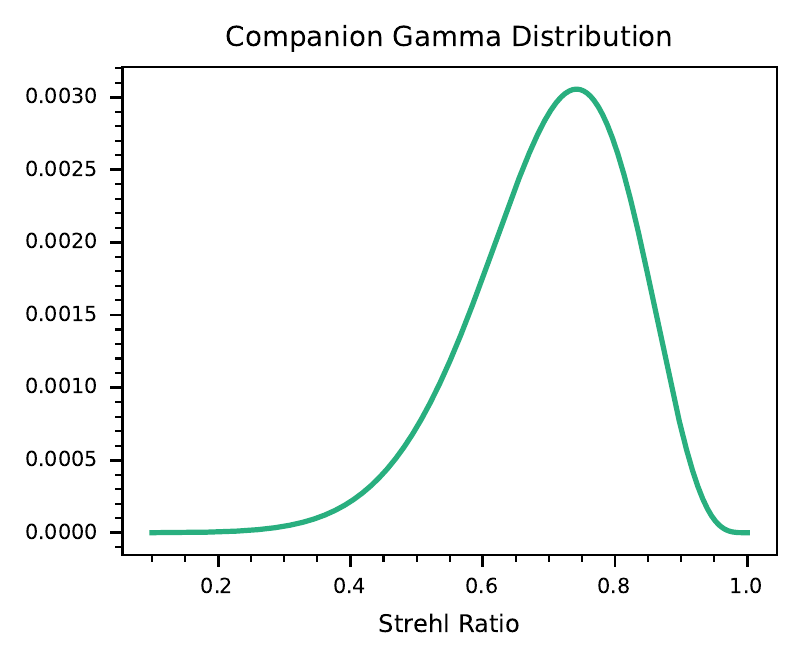}
    \caption{Gamma distribution (Equation \ref{eq:gammapdf}) from which the companion intensities were sampled in \S \ref{sec:realsimpcssd} and Figures \ref{fig:600_40_comp} \& \ref{fig:40_comp_back}. Here $k$ = 5, and $\Theta$ = 30, and $\mu$ corresponds to the median strehl ratio used (0.7).}
    \label{fig:gamma_used}
\end{figure}

To verify that a limitation for performing PCSSD on on-sky data is the assumption that the companion intensity ($I_P$) is constant, we generated new mock photon lists following the procedure as described in \citet{walter2019} but with the notable exception that the companion intensities were sampled from the Gamma PDF. The exact distribution from which the companion intensities were sampled can be seen in Figure \ref{fig:gamma_used} -- here sensible values of $k$, $\Theta$, and $\mu$ were chosen to roughly match observed values. These companion photons were also assumed to be correlated in time with a decorrelation timescale ($\tau$) of 0.1 s - the same as that of the MR. A suite of companion separations and contrasts was tested ranging from 3.5 to 12.5 $\lambda/D$ and $5\cdot10^{-5}$ to $4\cdot10^{-4}$ respectively. A suite of brighter companions was also tested (with contrasts ranging from $7.5\cdot10^{-4}$ to $6\cdot10^{-3}$) to more closely replicate the on-sky PCSSD results for binary stellar companions like HIP~109427 B which has a J-band contrast of $1.27\cdot10^{-3}$. 

\begin{figure*}
    \centering
    \includegraphics[width=0.9\textwidth]{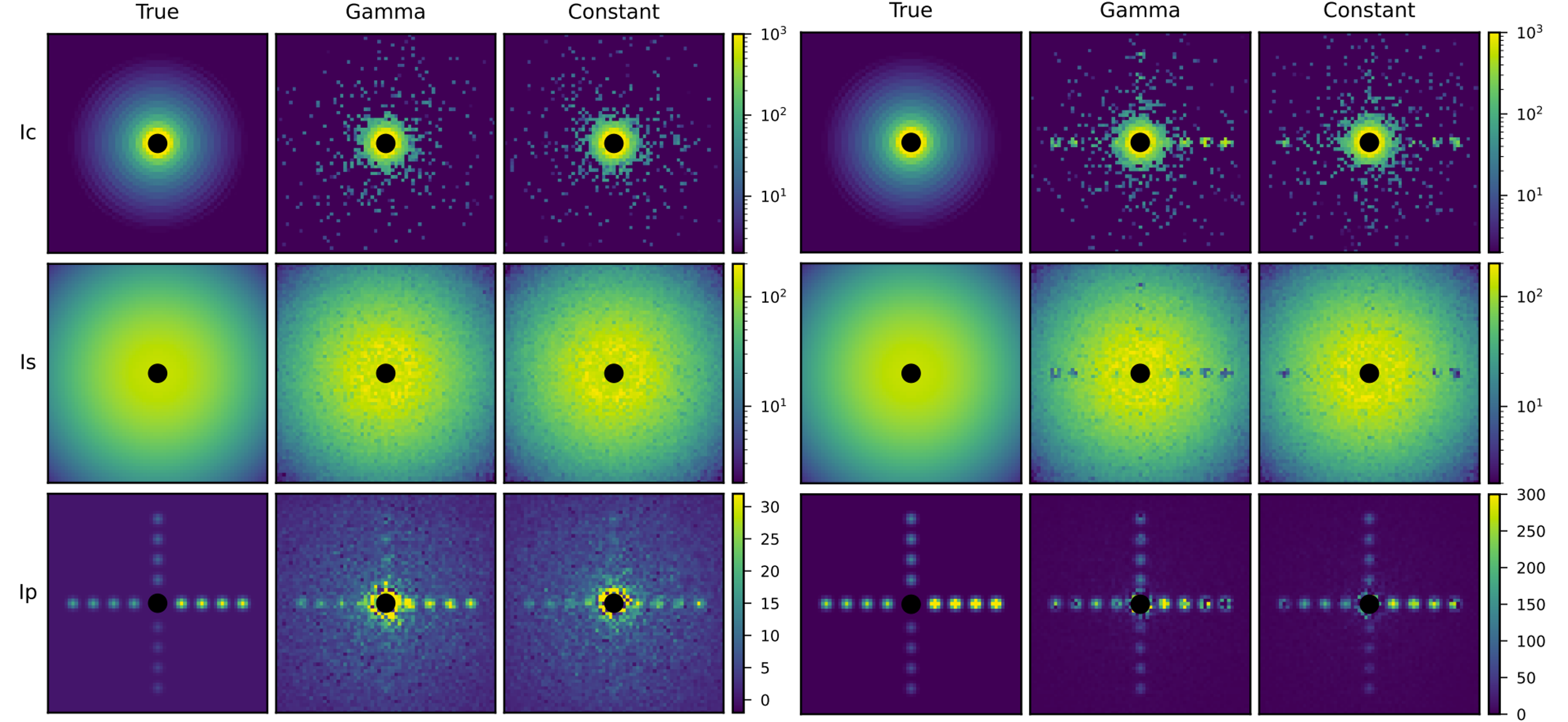}
    \caption{Simulated $I_C$, $I_S$ and $I_P$ images generated by running PCSSD on constant and Gamma distributed companion intensities for two different suites of companion contrasts. Left: $4\cdot10^{-4}$, $2\cdot10^{-4}$, $1\cdot10^{-4}$, $5\cdot10^{-5}$ Right: $6\cdot10^{-3}$, $3\cdot10^{-3}$, $1.5\cdot10^{-3}$, $7.5\cdot10^{-4}$ . The `True' columns are the input $I_C$, $I_S$, and $I_P$ images.} 
    \label{fig:600_40_comp}
\end{figure*}
The results are summarized in Figure \ref{fig:600_40_comp}. The addition of the Gamma distributed companion flux causes more companion light to be misattributed to $I_C$ over $I_P$ in the case of bright companions ($<1\cdot10^{-3}$ contrast - see right most column) matching observations.

Interestingly, in the case of higher contrast sources (left column) the shape of the companion distribution doesn't seem to make much of a difference in the performance of the PCSSD. We believe that in this regime there are not enough companion photons to shift the skewness of the underlying MR distribution at the location of companion pixels and so the light does not incorrectly end up in $I_C$. In other words, at these low count rates it becomes hard to distinguish between the Poisson and Gamma distributions. 

\begin{figure*}
    \centering
    \includegraphics[width=0.9\textwidth]{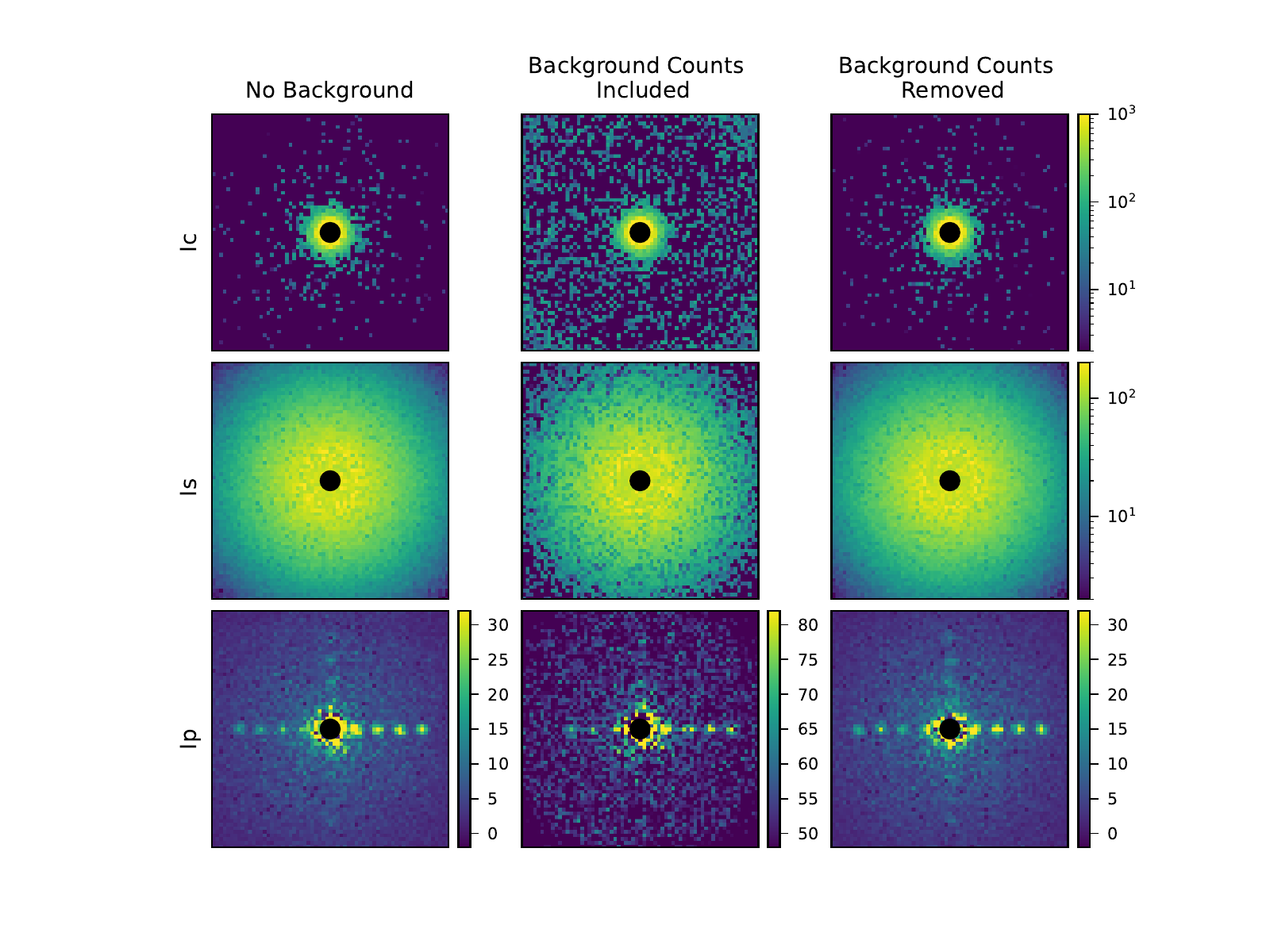}
    \caption{Simulated $I_C$, $I_S$ and $I_P$ images generated by running PCSSD on mock photonlists that sampled a Gamma PDF for the companion intensities. Left: No background included (same as the second column of Figure \ref{fig:600_40_comp}, left). Middle: Adding an additional constant and uncorrelated 50 photons/s/pixel background. Right: The 50 photons/s/pixel background is inserted, but then removed before running PCSSD. Any photons that were excluded from the analysis due to their proximity to a background photon (i.e.~falling within the 10 $\mu$s deadtime) are not accounted for which simulates performing a wavelength cut on MEC data in post-processing.}
    \label{fig:40_comp_back}
\end{figure*}

 At these higher contrasts however, other factors such as background noise sources are likely to become more significant. The effect of background counts on the performance of the PCSSD is summarized in Figure \ref{fig:40_comp_back}. Here the PCSSD was run on mock photon lists using the suite of higher contrast companions (see Figure \ref{fig:600_40_comp}, left panels) with an added constant and uniform 50 photons/s/pixel background with uncorrelated Poisson distributed arrival times. This count rate approximately matches the current IR background count rate for MEC and significantly degrades PCSSD performance (center column). 
 
 We also wanted to explore the effect of performing a wavelength cut on MEC data before running PCSSD. Since MEC stores the arrival time and wavelength of each incident photon, a wavelength range can be specified and only the arrival times of photons with wavelengths within that range returned. This is typically done in part to remove background counts since MEC's thermal background lies outside of our science bands (y-J). While these out-of-band photons themselves can be removed from analysis, their effect on temporally proximate photons still remains. Even if it is not used, an out-of-band photon incident on the detector still activates the 10 $\mu$s dead time of that pixel meaning that it is no longer photosensitive over that time range in a way that is unaccounted for by the PCSSD code. For the typical background count rates observed with MEC however this effect doesn't appear significant enough to impact PCSSD performance. This can be seen in the right column of Figure \ref{fig:40_comp_back} as the background removed $I_P$ images are consistent with the case of no thermal background present (left column). 

For all of these tests, the resulting $I_P$ signal-to-noise ratios (SNR) are summarized in Table \ref{snr}. SNR was calculated using $10^3$ independently generated 30s photon lists and is given by $\mathrm{SNR} = \left(\left<I_P\right> - \left<\mathrm{Background}\right>\right)/\left(\mathrm{std. dev.}\left<I_P\right>\right)$. Here the `Background' is determined using photon lists without any injected companions and `$\mathrm{std. dev.}\left<I_P\right>$' is the standard deviation of the mean companion intensity.

In summary, A constant 50 photons/s/pixel background significantly degrades the $I_P$ SNR of the faint companions at all separations when not removed. The perfect removal of these photons in post-processing recovers the results as if there had never been a background present, but this type of perfect subtraction can be challenging with real data due to the current energy resolution of MEC (R$\sim$5). An exploration of how imperfect background subtraction effects PCSSD performance will be left for future work. 

\begin{deluxetable*}{ccccccccccccc}
    \tabletypesize{\small}
    \tablecaption{Figure \ref{fig:40_comp_back} Companion Signal-to-Noise}
    \label{snr}
   \tablehead{\colhead{} & \multicolumn{3}{c}{Separation=3.5 $\lambda/D$} & \multicolumn{3}{c}{Separation=6.5 $\lambda/D$} & \multicolumn{3}{c}{Separation=9.5 $\lambda/D$} & \multicolumn{3}{c}{Separation=12.5 $\lambda/D$} \\ 
   \colhead{Contrast} & 
   \colhead{Control} & \colhead{50 cps}& \colhead{Removed}&
   \colhead{Control} & \colhead{50 cps}& \colhead{Removed}&
   \colhead{Control} & \colhead{50 cps}& \colhead{Removed}& \colhead{Control} & \colhead{50 cps}& \colhead{Removed}}
    \startdata
     $4\cdot10^{-4}$ & 4.5 & 3.4 & 4.2 & 6.3 & 4.4 & 6.8 & 7.4 & 3.0 & 7.0 & 5.8 & 1.3 & 6.4\\
     $2\cdot10^{-4}$ & 2.6 & 1.9 & 2.6 & 3.5 & 2.6 & 3.6 & 4.7 & 2.1 & 4.8 & 5.5 & 1.1 & 5.4\\
     $1\cdot10^{-4}$ & 1.6 & 1.0 & 1.7 & 2.2 & 1.4 & 2.2 & 2.6 & 1.5 & 2.7 & 3.4 & 0.8 & 3.6\\
     $5\cdot10^{-5}$ & 0.8 & 0.6 & 0.8 & 1.2 & 0.7 & 1.4 & 1.6 & 0.8 & 1.6 & 1.9 & 0.3 & 1.9\\
    \enddata
\end{deluxetable*}

\vspace{-0.8cm}
\section{Discussion}
To use photon arrival time statistics as a metric for differentiating photons from the bright speckle halo from those that originate from a faint companion, understanding and incorporating information about the distributions from which they originate is essential. This is especially true in regimes where the companion intensity is comparable to that of the underlying speckle field and deconvolving those two distributions becomes more important. 

Additionally, identifying and removing any possible sources of light outside of these distributions, or incorporating them explicitly into the model being used, is important in very high contrast regimes where background photons can easily outnumber photons coming from the faint source of interest.

Thankfully hardware fixes can remove out-of-band background light. Future work will identify and mitigate MEC's known IR background at wavelengths longer than its science band (y-J). This will significantly reduce the number of background counts and should aid the performance of even the current PCSSD algorithm on faint sources.

PCSSD has been derived only for a linear combination of a MR intensity distribution and a constant intensity component \citep{walter2019}.  Section \ref{sec:realsimpcssd} shows that a companion whose intensity instead follows a Gamma distribution can significantly weaken the sensitivity of PCSSD.  PCSSD could be expanded within its current analytical framework to account for a Gamma-distributed intensity component.  Doing so could require fitting an additional three parameters, rather than one for a constant-intensity companion, unless some of the Gamma distribution's parameters are independently known (e.g.~from AO telemetry). This could make performing the fit untenable for even typical MKID data sets that can easily run in excess of $10^9$ photons. 

Non-analytic approaches can be computationally less expensive than PCSSD and assume no knowledge about the underlying intensity distributions except for the fact that they are measurably different.  This work, and the works cited herein, have already demonstrated that intensity distributions differ for speckles and incoherent on-axis sources. The most prevalent non-analytic SSD techniques overlap with machine learning based approaches for detecting patterns in time series data such as Long Short-Term Memory recurrent neural networks \citep[LSTM;][]{hochreiter1997long}.
To facilitate further development of machine-learning-based SSD, the MKID Exoplanet Direct Imaging Simulator (MEDIS) can be used to generate more realistic simulated MKID data sets.
\citep{dodkins2020}.   

In high-contrast imaging there is unique information available at millisecond and microsecond timescales that is lost when taking long exposure images. New instruments deploying fast, noiseless detectors (like MEC) are now able to access this information and use it to start suppressing/differentiating quasi-static speckles and companions. Though not discussed here, work using millisecond images (sometimes combined with wavefront sensor telemetry) to directly measure and remove quasi-static speckles on-sky is another promising path forward. However, these techniques have additional challenges needing to run in conjunction with the AO loop during observing instead of in post-processing \citep{Martinache2014, Gerard2018, rodack2021}. Moving forward, the advancement of both real-time and post-processing techniques that leverage the information available at these fast timescales will be necessary for reaching the best achievable contrasts with current and next generation telescopes. 

\vspace{-0.75cm}
\section{Acknowledgements}
S.S. is supported by a grant from the Heising-Simons Foundation. T.B. gratefully acknowledges support from the Heising-Simons foundation and from NASA under grant \#80NSSC18K0439. N.Sk. and V.D. acknowledge support from NASA grant \#80NSSC19K0336. K.A. acknowledges funding from the Heising-Simons foundation.

The development of SCExAO was supported by the Japan Society for the Promotion of Science (Grant-in-Aid for Research \#23340051, \#26220704, \#23103002, \#19H00703 \& \#19H00695), the Astrobiology Center of the National Institutes of Natural Sciences, Japan, the Mt Cuba Foundation and the director’s contingency fund at Subaru Telescope. The authors wish to recognize and acknowledge the very significant cultural role and reverence that the summit of Maunakea has always had within the indigenous Hawaiian community, and are most fortunate to have the opportunity to conduct observations from this mountain.

\facilities{Subaru (SCExAO \citep{Jovanovic2015} + MEC \citep{Walter2020})}

\software{The MKID Pipeline \citep{steiger2022}, Matplotlib \citep{Huntermatplotlib}, SciPy \citep{2020SciPy-NMeth}, NumPy \citep{harris2020numpy}}
\bibliography{photon_stats}{}
\bibliographystyle{aasjournal}

\end{document}